# Optimization decision model of vegetable stock and pricing based on TCN-Attention and genetic algorithm

Xia, Linhan*

Beijing normal university-Hong Kong Baptist university united international college, temp.xialinhan@gmail.com

Zhang, Jinyuan

Beijing normal university-Hong Kong Baptist university united international college, m18083607612@163.com

Wen, Bohan

Beijing normal university-Hong Kong Baptist university united international college, q030026159@mail.uic.edu.cn

With the expansion of operational scale of supermarkets in China, the vegetable market has grown considerably. The decision-making related to procurement costs and allocation quantities of vegetables has become a pivotal factor in determining the profitability of supermarkets. This paper analyzes the relationship between pricing and allocation faced by supermarkets in vegetable operations. Optimization algorithms are employed to determine replenishment and pricing strategies. Linear regression is utilized to model the historical data of various products, establishing the relationship between sale prices and sales volumes for 61 products. By integrating historical data on vegetable costs with time information based on the 24 solar terms, a cost prediction model is trained using TCN-Attention. The Topis evaluation model identifies the 32 most market-demanded products. A genetic algorithm is then used to search for the globally optimized vegetable product allocation-pricing decision.

## 1 INTRODUCTION

With the rapid development of modern socioeconomics and the evolving lifestyles of consumers, supermarkets and shopping malls have gradually become the primary venues for urban residents to purchase daily vegetables. Alongside the proliferation of large-scale supermarkets, China's vegetable market has also undergone swift expansion. However, this growth has ushered in new challenges, particularly concerning the decisions on vegetable procurement costs and distribution quantities, which have emerged as pivotal factors determining the profitability of supermarkets.

In operational decision-making, pricing and allocation strategies play a crucial role in a merchant's profitability and market competitive standing. An appealing pricing strategy can boost consumer footfall, while an appropriate allocation ensures a balance between supply and demand. Yet, devising optimal strategies amidst numerous vegetable varieties remains a significant challenge.

Traditional strategies often rely on experience or oversimplified forecasting models, potentially failing to accurately capture the intricate dynamics of the market. Fortunately, advancements in big data and artificial intelligence offer more precise tools for tackling this issue. This paper leverages these cutting-edge technologies, particularly the TCN-Attention neural network and genetic algorithms, to study and analyze supermarkets' pricing and allocation strategies in vegetable operations, thereby providing more scientifically rigorous and efficient decision support.

*Correspondence: temp.xialinhan@gmail.com

## 2 RELATED WORK

Over the past several decades, pricing and distribution strategies in the vegetable market have been central topics in supply chain management andOver the past several decades, pricing and distribution strategies in the vegetable market have been central topics in supply chain management and retail research. Different scholars have proposed various methods to advance studies in this area, especially with the application of artificial intelligence and algorithm optimization techniques.

### 2.1 Statistical Forecasting Methods

In early research, statistical forecasting methods were predominant. For instance, moving averages and exponential smoothing [1] provided a preliminary framework for market predictions. These approaches employ simple calculations based on historical data to anticipate future market demand and prices. However, they might face accuracy and stability issues when dealing with intricate market dynamics.

### 2.2 Machine Learning-based Forecasting

As machine learning technology advanced, algorithms such as decision trees, random forests, and support vector machines started gaining attention. For example, J Huber and his team simulated and predicted market demand and price fluctuations using these methods and validated their effectiveness in various application scenarios [2].

### 2.3 Deep Learning and Time Series Forecasting

In recent years, advanced deep learning algorithms have been used for time series predictions. Recurrent neural networks such as LSTM and GRU have become popular because of their exceptional performance in sequence data analysis [3]. They effectively capture long-term and short-term dependencies in time series data using memory structures.

### 2.4 Temporal Convolutional Neural Networks (TCN)

TCN offers a fresh perspective on time series analysis. Unlike traditional recurrent networks, TCN employs convolutional structures to process data across different time scales, effectively capturing long-term dependencies [4]. In some applications, its performance even surpasses that of LSTM and GRU.

### 2.5 Introduction of Attention Mechanism

The attention mechanism, proposed by Vaswani A and others, has brought about a revolutionary change in deep learning models [5]. It assists models in automatically assigning varying weights to data points of different importance within a time series. In pricing strategies of the vegetable market, this method helps the model concentrate on data features crucial for pricing decisions, thus enhancing prediction accuracy.

In summary, despite the extensive study of pricing and distribution strategies in the vegetable market, the advancement of technologies, particularly the applications of big data and artificial intelligence, ensures numerous research opportunities in this field. This paper's research builds on previous studies, combined with the latest technological advancements, to provide a more comprehensive and efficient solution.

## 3 COST PREDICTION MODEL

When determining the pricing and distribution strategies for each vegetable, precise cost forecasting is essential. Given that the wholesale prices of vegetables are significantly influenced by seasons and time [6], this paper incorporates the traditional Chinese 24 solar terms into the prediction model. Alongside time series, this research predicts the cost for each product over the next seven days. We employ the TCN-Attention fusion model for forecasting, which supports multi-input and single-output.

### 3.1 Data preprocess

**Hierarchical One-Hot Encoding for the 24 Solar Terms:** To precisely capture the influence of the traditional Chinese 24 solar terms on vegetable costs, this paper employs a unique hierarchical one-hot encoding method. Direct numerical encoding might overlook the periodic characteristics of the solar terms, while conventional one-hot encoding can produce high-dimensional sparse data, both of which could impair model performance. We adopt hierarchical one-hot encoding, taking into account the relationship between seasons and the solar terms. Seasons are initially one-hot encoded, followed by further encoding based on the sequence of the six solar terms within each season. Combining the encoding results from both stages provides the hierarchical encoding for the 24 solar terms, as illustrated in Table 1.

Table 1 Encoding result

| Solar term | Hierarchical One-Hot Encoding for the 24 Solar Terms |
|---|---|
| Li Chun | [1,0,0,0,1,0,0,0,0,0] |
| Qing Ming | [1,0,0,0,0,0,0,0,1,0] |
| Da Han | [0,0,0,1,0,0,0,0,0,1] |

Using hierarchical One-hot coding, the dimension of solar term information training data is reduced by 53.8% compared with One-hot coding, which significantly improves the information density and expression ability of the coding results.

Time Series Sliding Window[7]: To forecast the cost of a product over the next seven days, this paper employs the sliding window approach for data preprocessing. This method segments data into fixed-length windows, extracting crucial patterns and information from historical data, as illustrated in Figure 1.

Window Configuration: Each window encompasses 15 days of data, implying that the model leverages the prior 15 days of data to predict the subsequent 7-day cost.

Solar Term Information: The model not only factors in the cost data from the preceding 15 days but also incorporates the solar term information for the upcoming 7 days as an additional feature.

Window Shift: The window slides with a step size of one day, generating new data subsets for training.

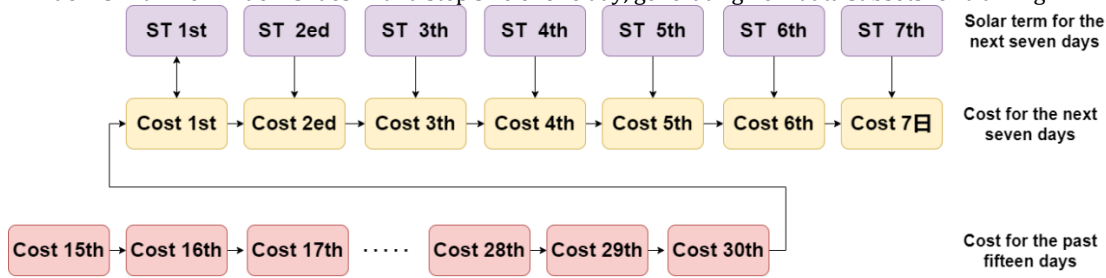

Fig 1 Sliding window processing flow

With a sliding window, we get observed variables over a small time period to better capture the dynamics and seasonality of product prices, which allows the model to learn these short-term dependencies and make more accurate predictions accordingly.

**Data normalization**: After the 24 solar terms coding and sliding window are completed, all data should be normalized. The normalization algorithm used in this paper is MinMaxScaler algorithm, as shown in Formula 1.

$$x_{scaled} = \frac{x - x_{min}}{x_{max} - x_{min}} \tag{1}$$

### 3.2 Model structure

The Model constructed in this paper is the temporal convolutional neural network -Attention mechanism feature fusion Model (TCN-Attention-feature fusion model), which is composed of two input layers, a feature fusion layer, and a fully connected output layer, as shown in Figure 2 below.

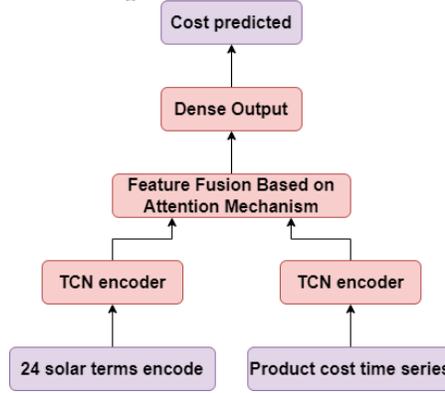

Fig 2 Model structure

**Temporal convolutional layers:**

The input layer of the model comprises two Temporal Convolutional Networks (TCNs)[8], separately processing the encoding of the 24 solar terms and the cost time series of the product categories. These two TCN layers optimize the model's fitting accuracy and enhance its time series inference capabilities, as depicted in Figure 3.

Notably, the TCN employs causal convolution[9], which is a convolution operation ensuring that the output at the current timestep is solely based on the current and previous information, unaffected by future data. This guarantees a clear causal relationship, where the output is entirely defined by the input and its historical information. The mathematical representation can be seen in formula 2.

$$y[t] = \sum_{r=0}^{K} w[r] \cdot x[t - r] \tag{2}$$

Where y[t] is a given time series, w[r] is the value of the convolution kernel at time r, and K is the length of the convolution kernel. Causal convolution also limits the model's ability to obtain information from future data, improving the model's reasoning ability.

In contrast to traditional sequential neural networks such as LSTM, the TCN input layer used in this model is characterized by dilated convolutions. The difference lies in the discontinuity of the convolution kernel on the time series, operating at large intervals. See Formula 3 for details.

$$y[t] = \sum_{r=0}^{K} w[r] \cdot x[t - r \cdot d] \tag{3}$$

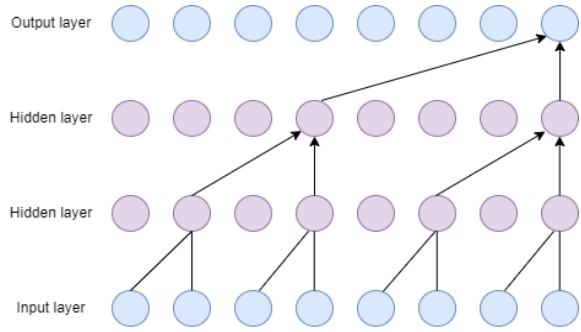

Fig 3 TCN Neural Network model

**Feature Fusion Based on Attention Mechanism:**

The model adopts a multi-input single-output architecture, necessitating feature fusion during both the training and prediction phases. The attention mechanism[5] has been widely utilized in models with multiple information sources or multi-modalities, as in this model. It receives two types of inputs: the first being the cost time series of vegetable categories processed by the TCN, and the second is the hierarchical one-hot encoding of the 24 solar terms' time series, also processed by the TCN. To accurately fuse these two types of information, this study introduces a feature fusion layer based on the attention mechanism, as illustrated in Figure 4.

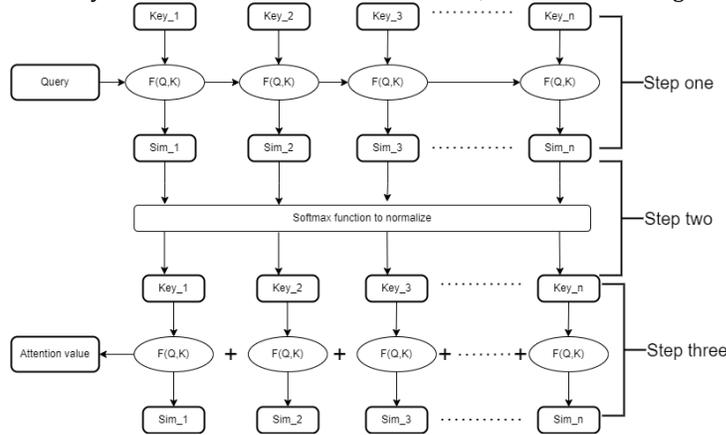

Fig 4 Attention Mechanism

The attention mechanism automatically focuses on important information in the neural network to realize dynamic adjustment and enhance the interpretability of the model. In this paper, 24 solar terms are used as key values, historical cost time series are used as query values, and future cost time series are output for attention operation. The operation process is as follows:

In the first step, the two feature matrices $x_1, x_2$ are concatenated horizontally to compute the weights for each query and each Key:

$$K = Concat(x_1, x_2) \tag{4}$$

$$Sim(Q, K_i^T) = Q \cdot K_i^T \tag{5}$$

In the second step, Softmax is used to normalize the weights to obtain the attention weights for each feature:

$$\alpha_i = Softmax(e^{Sim_i}) = \frac{e^{Sim_i}}{\sum_{j=1}^{L_x} e^{Sim_i}} \quad (6)$$

In the third step, with the attention weights, we can compute the weighted average of the fused features:

$$Fused\ Features = A(Q, K, V) = \sum_{i=1}^{L_x} \alpha_i \cdot V_i \quad (7)$$

After fusing the features, the final prediction is made through a fully connected output layer. The purpose of this layer is to map the fused features to specific values for cost prediction.

$$Predicted\ Cost = Fused\ Features \cdot W_o + b_o \quad (8)$$

The model is obtained by training the model on all vegetable products separately, so as to predict the specific unit cost of vegetable products in the next seven days.

## 4  SALES FORECASTING MODEL

To precisely formulate restocking and pricing strategies, this paper aims to align the forecasted restocking quantity of vegetable categories with the actual sales volume of the upcoming week. Based on this, the paper adopts a time series model based on TCN-Attention and Bootstrap[10] to predict the sales volume range for the next seven days, serving as a constraint for strategy optimization. For each category, the study trains 100 independent models, each based on a continuous subset of the normalized data, ensuring the continuity of the time series. By training models for 61 vegetable products, a confidence interval prediction for future sales is obtained, which follows a normal distribution (as shown in Figure 5).

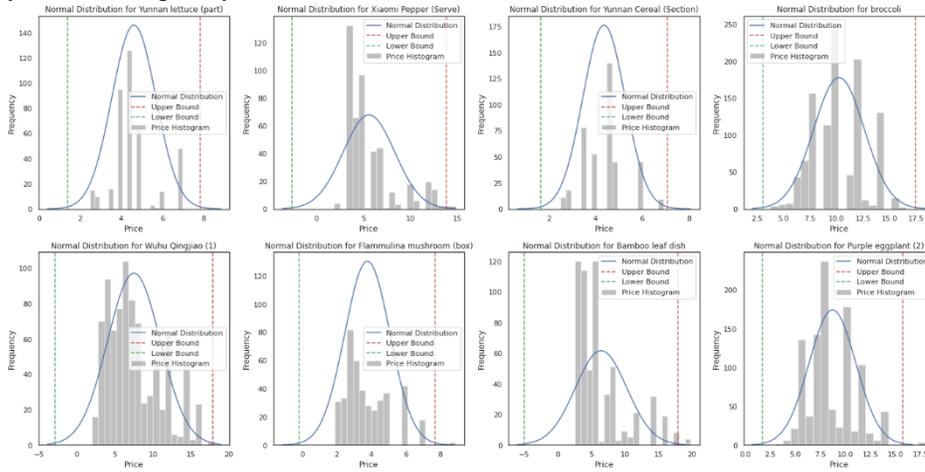

Fig 5 Product sales range (partial)

## 5 GENETIC ALGORITHM OPTIMIZATION MODEL

### 5.1 TOPSIS evaluation model

To cater to market demands and ensure maximal vegetable sales profit in supermarkets, this paper, grounded on market supply and demand theory, treats the demand volume as sales volume, integrating it into the scoring factors. Moreover, the profit and discount rate of goods play a crucial role in maximizing returns. Thus, core indicators for evaluating the suitability of vegetable sales encompass sales volume and profit. This study adopts the entropy method to assign weights and employs the TOPSIS[11] method to score each product, thereby determining the optimal sales mix.

Initially, the weights were computed using the entropy method. This multi-criteria decision-making method aims to establish weights for different indicators, offering a better representation of each indicator's importance in the evaluation process. As the entropy method is objective, it minimizes biases stemming from subjective weight assignment. Before evaluation, it is essential to normalize and standardize the data for evaluation indicators, ensuring non-negativity.

$$z_{ij} = \frac{x_{ij} - x_{min}}{x_{max} - x_{min}} \tag{9}$$

Where $x_{ij}$ corresponds to the evaluation index data, $min$ and $max$ represent the maximum and minimum values in the data, respectively, and $z_{ij}$ represents the index data that has been processed. For the processed data $z_{ij}$, the corresponding probability value is:

$$p_{ij} = \frac{z_{ij}}{\sum_{i=i}^{n} z_{ij}} \tag{10}$$

According to the definition of information entropy, for a specific index, the entropy value can be used to judge its dispersion degree. The smaller the information entropy is, the greater the dispersion degree of the index is, and the greater the weight it takes in the evaluation index. To solve the entropy value of the $jth$ vegetable product, the information utility entropy $\hat{d}_j$ needs to be calculated to measure its information amount.

$$e_j = -\frac{1}{\ln n} \sum_{i=1}^{n} p_{ij} \ln(p_{ij}) \tag{11}$$

$$\hat{d}_j = 1 - e_j \tag{12}$$

Finally, $\hat{d}_j$ is normalized to obtain the weight $w_j$ corresponding to each indicator type, see formula. 13.

$$w_j = \frac{d_j}{\sum_{j=1}^{m} d_j} \tag{13}$$

The corresponding weights are shown in Table 2.

Table 2 Entropy weight

| Metrics | Information entropy value | Information utility value | Weight (%) |
|---|---|---|---|
| Total profit | 0.957 | 0.043 | 19.888 |

| Metrics | Information entropy value | Information utility value | Weight (%) |
|---|---|---|---|
| Total sales volume | 0.827 | 0.043 | 80.112 |

After the entropy weight calculation is completed, the Topsis vegetable single item score is carried out. TOPSIS is a multi-criteria decision making method used to evaluate the overall score of an item based on the weights of the indicators and the characteristics of the item. The method is based on the concepts of ideal and negative ideal solutions, which represent the best and worst performance cases, respectively. The composite score of a commodity depends on the distance between it and the ideal and negative ideal solutions, as well as the weights.

The $z_i$ values with the largest and smallest data in each index are found to form vectors $Z^+$ and $Z^-$, which represent the best and worst vegetable index data respectively.

$$Z^+ = \{z_1^+, z_2^+, \ldots, z_3^+\} \tag{14}$$

$$Z^- = \{z_1^-, z_2^-, \ldots, z_3^-\} \tag{15}$$

For the $ith$ vegetable to the optimal data value, the distance of the worst data value is denoted as $D_i^+$ and $D_i^-$ and the formula is given in formula. 16, 17.

$$D_i^+ = \sqrt{\sum_{j=1}^{m}(z_i^+ - z_{ij})^2} \tag{16}$$

$$D_i^- = \sqrt{\sum_{j=1}^{m}(z_i^- - z_{ij})^2} \tag{17}$$

The score of specific vegetable i can be obtained by formula 18, $D_i^+$ and $D_i^-$ are distance values, and the value range of score $G_i$ is between [0,1], and the larger $G_i$ is, the smaller $D_i^+$ is, the closer to the ideal value, and the higher the score of vegetable products.

$$G_i = \frac{D_i^-}{D_i^+ - D_i^-} \tag{18}$$

By TOPSIS model, 32 vegetable products are selected as the optimized objects for the next model (see Table 3).

Table 3 Vegetable Product Rankings (Part)

| Product name | Ranking | Product name | Ranking |
|---|---|---|---|
| Yunnan lettuce | 1 | Cordyceps sinensis flower | 41 |
| Millet pepper | 2 | Red pepper | 42 |
| Yunnan cole wheat vegetables | 3 | Yunnan lettuce1 | 43 |
| Broccoli | 4 | Colorful pepper (2) | 44 |

| Product name | Ranking | Product name | Ranking |
|---|---|---|---|
| Wuhu green pepper (1) | 5 | Yunnan cole wheat vegetables1 | 45 |
| Flammulina | 6 | Honghu lotus root belt | 46 |
| Bamboo leaf vegetable | 7 | Tall melon (1) | 47 |
| Purple eggplant (2) | 8 | White Jade Mushroom | 48 |
| Screw pepper | 9 | Tall melon (2) | 49 |
| lettuce | 10 | Wood fungus | 50 |

### 5.1 Genetic algorithm optimization

For 32 vegetable products selected by TOPSIS model, this paper uses genetic algorithm to achieve the maximum benefit optimization under certain restrictions (see Table 4). Genetic algorithm is an optimization and selection algorithm, which is similar to the process of population evolution in nature. The model obtains the global optimal solution through heuristic search, crossover and mutation in continuous iteration.

Table 4 Genetic algorithm optimization constraints

| Constraints | |
|---|---|
| 1. | The sales volume is within the prediction interval of the Bootstrap sampling model mentioned above |
| 2. | The pricing and distribution quantities must be greater than zero |

In order to maintain the diversity of population and provide continuous search, Gaussian mutation is used as the mutation algorithm. The Gaussian mutation algorithm mutates one or more gene positions on a chromosome. At the selected gene location, a randomly drawn value from a normal (Gaussian) distribution is added as a perturbation. By adjusting the standard deviation of the Gaussian distribution, the strength of the variation can be controlled. For example, if we think that the current search is close to the optimal solution, we can reduce the standard deviation to perform a more refined local search. Conversely, if a large range of exploration is deemed necessary, the standard deviation can be increased to increase the strength of the variation (see Equation 19).

$$f(x|\mu, \sigma^2) = \frac{1}{\sqrt{2\pi\sigma^2}} e^{-\frac{(x-\mu)^2}{2\sigma^2}} \quad (19)$$

The pseudocode of the GA combined in this paper is as follows.

ALGORITHM 1: GA in optimization

```
INITIALIZE population with n individuals
FOR EACH individual in population
    CALCULATE its fitness
END FOR

FOR gen FROM 1 TO ngen
```

```
   SELECT parents for reproduction
   PERFORM crossover on parents to produce offspring
   PERFORM mutation on offspring
   FOR EACH individual in offspring
      REPAIR the individual to keep values within constraints
      CALCULATE its fitness
   END FOR
   SELECT individuals for the next generation from parents and offspring
   STORE statistical info (max, min, average) for this generation
END FOR

DISPLAY the statistical info graphically
SELECT and PRINT the best individual from the last generation
```

The optimal distribution volume-pricing decision is obtained through calculation and iteration(see figure 6).

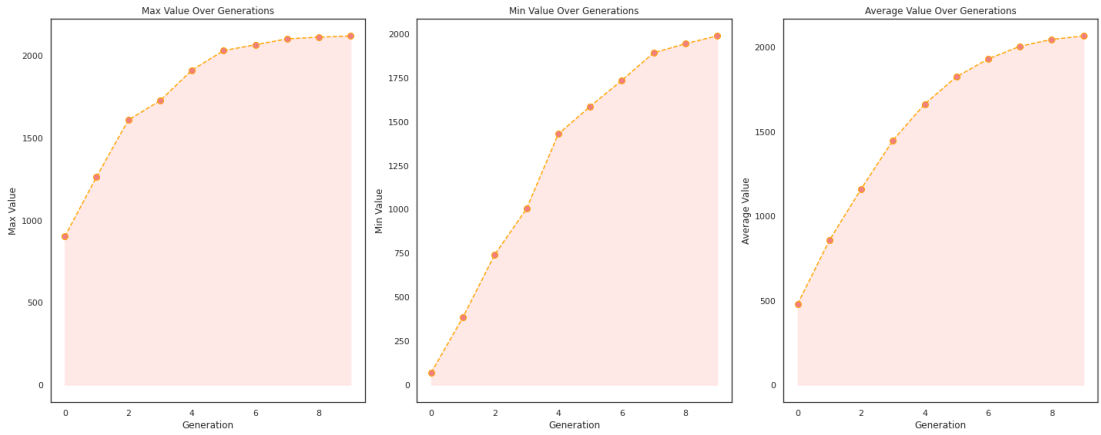

Fig 6 Genetic algorithm optimization iteration

## 6  EXPERIMENTS AND RESULTS CONCLUSION

This paper proposes a cost prediction model based on TCN-Attention feature fusion, a sales interval prediction model and a revenue maximization optimization model based on genetic algorithm, so as to realize the pricing-distribution decision model. In the experiment, the cost prediction model and the revenue maximization optimization model are tested and analyzed.

### 6.1 TCN-Attention model

For the TCN-attention cost prediction model, this paper compares LSTM, TCN, RNN, and the three indicators of this model MSE, MAE and RMSE are shown in Equations 20,21,22. The experimental results are shown in Table 5.

$$MSE = \frac{1}{n}\sum_{i=1}^{n}(y_i - \hat{y_i})^2 \qquad (20)$$

$$MAE = \frac{1}{n}\sum_{i=1}^{n}|y_i - \hat{y_i}| \tag{21}$$

$$RMSE = \sqrt{MSE} \tag{22}$$

Table 5 Comparison of experimental results

| Model | MSE | MAE | RMSE |
| --- | --- | --- | --- |
| TCN-Attention | 0.043 | 0.113 | 0.2074 |
| TCN | 0.067 | 0.134 | 0.2588 |
| LSTM | 0.092 | 0.252 | 0.3033 |
| **RNN** | **0.353** | **0.3473** | **0.5941** |

Experiments show that the TCN-Attention model combining the 24 solar terms in traditional Chinese culture as seasonal and time features is superior to other time series prediction models in the three regression evaluation indicators of MSE, MAE and RMSE.

### 6.2 Based on genetic algorithm revenue maximization optimization model

The genetic algorithm is used in this paper to optimize the distribution volume and the pricing of vegetable products. In this paper, the average daily income is compared with the average daily income from July 1 to 7, 2023, and the results are shown in Table 6.

Table 6 Comparison of average daily return and true return based on the optimized strategy

|  | Average daily return |
| --- | --- |
| True earnings from July 1-7 | 723.7 |
| Optimize revenue based on genetic algorithm | 2107.9 |

By comparing the real profit and the profit of the optimal decision model based on genetic algorithm, the decision-making model proposed in this paper significantly improves the average daily profit of the supermarket in vegetable products.

In this paper, the TCN-Attention model is combined with the 24 solar terms information to predict the daily cost of a single vegetable product, and the genetic algorithm and TOPSIS model are used to select and optimize the types of vegetables, distribution and pricing, so as to realize the optimization of supermarket in the sales of vegetable products. In this paper, time series factors are considered in the prediction of purchasing cost, the factors of different seasons are introduced, and the influence of date on vegetable crops according to 24 solar terms is taken into account. At the same time, the entropy weight method can objectively reflect the importance of each index in the evaluation process and avoid the error caused by subjective weighting.

At the same time, there are still some shortcomings in this study. Among them, the index selection of TOPSIS cannot determine the appropriate number, and the key indicators need to be selected by personal analysis and evaluation. This can be optimized in subsequent research.